\documentclass[twocolumn,showpacs,preprintnumbers,amsmath,amssymb]{revtex4}

\usepackage{graphicx}
\usepackage{dcolumn}
\usepackage{bm}
\usepackage{url}

\begin{document}

\title{Single Superconducting Split-Ring Resonator Electrodynamics}
\author{Michael C. Ricci}
\author{Steven M. Anlage\footnote[0$^{a)}$]{$^{a)}$email: anlage@squid.umd.edu}}
\affiliation{Center for Superconductivity Research, Department of Physics, University of Maryland, College Park, MD 20742-4111}

\date{\today}

\begin{abstract}
We investigate the microwave electrodynamic properties of a single superconducting thin film split-ring resonator (SRR).  The experiments were performed in an all-Nb waveguide, with Nb wires and Nb SRRs.  Transmission data showed a high-Q stopband for a single Nb SRR ($Q \sim 4.5\times10^4$ at 4.2 K) below $T_c$, and no such feature for a Cu SRR, or closed Nb loops, of similar dimensions.  Adding SRRs increased the bandwidth, but decreased the insertion loss of the features.  Placing the Nb SRR into an array of wires produced a single, elementary negative-index passband ($Q \sim 2.26\times10^4$ at 4.2 K).  Changes in the features due to the superconducting kinetic inductance were observed.  Models for the SRR permeability, and the wire dielectric response, were used to fit the data.
\end{abstract}


\maketitle

	Since the end of the last century, many researchers have been interested in verifying predictions by Veselago \cite{veselago:517} for materials with a simultaneously negative permittivity ($\epsilon$) and permeability ($\mu$).  Such materials have a negative index of refraction ($n < 0$), at least within a limited frequency range.  Many of these experiments have been performed at room temperature with thick normal metal split-ring resonators (SRRs) and wires, often prepared on lossy dielectric circuit board \cite{shelby:77,bayindir:120,houck:137401}.  Experiments utilizing photonic crystals \cite{cubukcu:604} demonstrate behavior similar to that expected from a medium with $n < 0$.  A recent experiment demonstrated a negative-index passband at cryogenic temperatures with superconducting metals, thin films, and low loss dielectrics \cite{ricci:034102}. 
	
	A predicted property of negative index of refraction materials is the ability to focus with a flat lens \cite{veselago:517}, and amplify evanescent waves, leading to super-resolution imaging \cite{pendry:3966}.  However, amplification is partly limited by losses, such that only a limited range of wavenumbers is amplified \cite{shen:3286,gomez-santos:077401,smith:1506}.  In Ref. [6], it was proposed that a superconductor can attain the low losses needed to achieve significant evanescent wave amplification at microwave frequencies.  In addition, applications such as improved radiating properties of ultra-small dipole antennas \cite{ziolkowski:2626}, require reducing the dimensions of existing metamaterials at $\sim 10$ GHz by at least an order of magnitude.  However, the normal-metal losses in $\mu_\text{eff}$ of a thin film SRR scale as \cite{dimmock:2397,zhou:223902} $\Im[\mu_\text{eff}] \sim \rho/d \ell$, where $\rho$ is the resistivity, $d$ the SRR film thickness, and $\ell$ the lattice parameter.  Similarly, wire losses scale as \cite{pendry:4773} $\Im[\epsilon_\text{eff}] \sim \rho/r^2$, where $r$ is the wire radius.  Thus, normal-metal metamaterials will become increasingly lossy as their size, thickness, and spacing are reduced, destroying their properties of $\epsilon_\text{eff}$, $\mu_\text{eff}$, and $n_\text{eff} < 0$.  
	
	In Ref. [6], it was shown that due to strong interactions, the superconducting multiple-SRR array had a jagged transmission spectrum, rather than a single, smooth transmission minimum.  Each dip was assumed to be due to the $\mu_\text{eff}<0$ behavior of individual SRRs, or a small number of SRRs with degenerate frequency.  Harvesting a single SRR from the array should produce a single dip, and the ability to experiment on a single SRR would greatly simplify the problem of understanding the collective behavior.  Thus, a series of experiments were performed on a single Nb SRR.  This Letter also shows that in order to scale down SRR dimensions and remain in the microwave regime it is necessary to use superconducting thin films, which have the high quality factor ($Q$) needed to produce a $\mu_\text{eff}<0$ medium.

	An all-Nb X-band (WR90) waveguide (cross-section $22.86 \times 10.16$ mm$^2$), with a critical temperature $T_c$ = 9.25 K (as measured by AC susceptibility), was coupled to an Agilent 8722D vector network analyzer via 1.37 m of Cu coaxial cable, and all-Nb waveguide-to-coax couplers with Cu antennas.  The waveguide was perforated with 0.51-mm-diameter holes through the broad sides, spaced 4.57 mm from each other and 2.29 mm from the walls, allowing for 5 vertical wires (parallel to the applied electric field, $\mathbf{E}$) across the width, and 27 along the length.  A single square Nb SRR (details in Ref. [6]) was embedded in a thin slab of Rohacell 51 ($\epsilon_r = 1.07$ at 10 GHz), and inserted into the center of the waveguide.  To compare the results with normal-metal metamaterials, an SRR made of copper/nickel/gold (35 $\mu$m/30 $\mu$m/2 $\mu$m thick) on a 0.80-mm-thick G10 substrate was also tested.  This SRR was slightly larger than the Nb one; it had an outer loop side length of 2.6 mm, inner loop side length of 1.6 mm, gap width of 0.3 mm, and a line width of 0.2 mm.  (A 216-SRR array of these SRRs resonated between 9 and 10 GHz in a Cu X-band waveguide.)  All experiments were performed between 4.2 and 10 K in vacuum, with a temperature control of $\pm 5$ mK.  Closed Nb loops of similar dimension to the Nb SRR were tested as a control, since the closed loops lack the resonant magnetic response in the frequency range of interest \cite{koschny:107402}, and have no bianisotropy \cite{padilla:0508307}.

	Fig. 1 shows the results of a microwave transmission experiment for a single Nb SRR in the waveguide, at frequencies above the waveguide cutoff frequency (6.5 GHz).  The single SRR was oriented to couple to the magnetic field of the TE$_{10}$ mode, with the imaginary line completing the capacitive gaps orthogonal to $\mathbf{E}$ (top left inset).  A single dip in transmission was observed in the superconducting state, where the dip is consistent with a frequency region of $\epsilon_\text{eff} > 0$ and $\mu_{\text{eff}} < 0$.  This result, along with the observation of a transmission peak near the same frequency when the Nb SRR is embedded in an $\epsilon_\text{eff}<0$ medium (presented below), allows us to exclude the possibility of an $\epsilon_\text{eff}<0$ feature producing this dip due to bianisotropy \cite{marques:144440}.  This dip decreased in depth and widened as the temperature increased, due to the increasing surface resistance of the Nb.  Above the critical temperature ($T_c$ = 8.65 K by AC susceptibility), the dip disappeared, consistent with a large value of $\rho/d$ in the normal-metal state of the Nb film, which causes the $\mu_\text{eff} < 0$ region to disappear \cite{zhou:223902}.  We estimated the $Q$ of the Nb SRR resonant feature (defined by the 3 dB point above the minimum) at 4.2 K to be $4.5\times10^4$, which is about 62 times greater than the scaled value (by $(\omega_\text{0,Nb}/\omega_\text{0,NM})^{1/2}$) for the single normal-metal SRR in Ref. \cite{gay-balmaz:2929}, despite the Nb one being 250 times thinner, and with an area 162 times smaller.  The Nb SRR was also tested with the capacitive gaps parallel to $\mathbf{E}$ ({\it i.e.} coupling the gaps to {\bf E}), which resulted in an increase of the dip frequency by 30 MHz, but the $Q$ and insertion loss remained unchanged.  This demonstrates that the added electric polarization and magnetization induced through bianisotropy has a small net effect on the SRR resonant properties.  The Cu-Ni-Au SRR response was featureless ({\it i.e.}, $|\text{S}_{21}| \sim 1$) down to 4.2 K, consistent with the idea that $\rho/d$ is too large to produce a $\mu_\text{eff} < 0$ feature.  The closed Nb loop response was also featureless between 3 and 18 GHz.  All these results indicate that the SRR dip is consistent with a $\mu_\text{eff}<0$ response. 

\begin{figure}
\includegraphics[width = 2.7 in,angle=270]{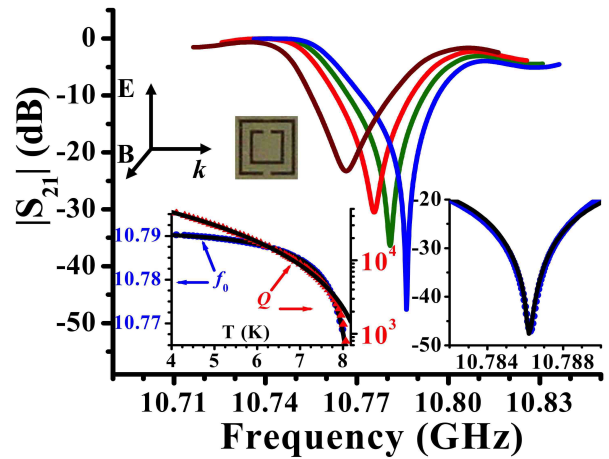}
\caption{Fig. 1 (Color Online) Transmission v. frequency for a single Nb SRR in an empty Nb X-Band waveguide for 5.05 (blue), 7.70 (green), 7.80 (red), and 8.10 (maroon) K (right to left).  Right inset is the 5.05 K dip (blue dots) and fit (solid line).  Bottom left inset is the frequency of minimum transmission (blue dots, left axis) and $Q$ (red triangles, right axis) with fits (black lines) v. temperature.  Top left inset is the SRR and incoming wave.  (Data corrected for cable loss.)}
\label{fig1}
\end{figure}

The resonant feature was modeled with a transfer matrix model \cite{baena:75116}, using a $\mu_{\text{eff}} \sim \mu(f_0, l, \Gamma)/\mu_0$ from Eq. (43) in Ref. \cite{pendry:2075}.  According to this model, the dip in $|\text{S}_{21}(f)|$ is mainly due to an impedance mismatch between the SRR-loaded section of the waveguide and the empty sections of the waveguide, and not due to absorption, consistent with the zero-loss results in Ref. \cite{pendry:2075}.  A nonlinear least squares algorithm fit the transmission data with parameters of loss ($\Gamma$), resonant frequency ($f_0$), material length ($l$), and magnetic filling fraction ($F$).  The fit (black line) is shown in the right inset of Fig. 1, and matched the 5.05 K data (blue dots) with $f_0 = 10.79$ GHz, while $l = 0.18$ mm, and $F = 0.13$, which are near the expected values of $l \sim 1$ mm, and $F \sim 0.1$.  As an indication of the low-loss dielectric and superconducting metals, $\Gamma/2\pi = 373$ kHz, which is smaller than the corresponding result for the Nb SRR array in Ref. [6], due to the use of a low-loss Nb waveguide, and the lack of interactions with other SRRs.  
	
Below $T_c$, the kinetic inductance of the superfluid electrons in the Nb film must be included in the total inductance of the SRR \cite{anlage:1388}.  As the temperature increases, the kinetic inductance will increase, thereby decreasing the resonant frequency.  This is shown in the data (blue circles, left axis) and fit (black line) of the bottom left inset of Fig. 1.  If we assume a fixed filling fraction, and approximate the SRR as a single-ring {\it LC} resonator, of effective radius $R$ and capacitance $C$, we can write the inductance as $L = L_{\text{geo}} + L_{\text{kin}}$, where $L_\text{geo} = \mu_0R\kappa$ is the geometrical inductance, and $\kappa \approx 4.332$ (for our geometry) can be found in Ref. \cite{hoffmann}.  The temperature dependent kinetic inductance can be approximately written as \cite{henkels:63,bluzer:2051} $L_\text{kin}(T) \approx 2\pi \mu_0 R w^{-1} \lambda(T)\coth(d/\lambda(T))$, with ring width $w$, thickness $d$, and two-fluid model penetration depth  $\lambda(T)$.  The resonant frequency, $2\pi f_0 = (LC)^{-1/2}$, should therefore behave as $f_0(T) = f_{0,g}[1 + \beta \xi^{-1}\coth(\chi \xi)]^{-1/2}$, where $2\pi f_{0,g} = (L_\text{geo}C)^{-1/2}$, $\beta = 2\pi\lambda(0)/(\kappa w) \sim 10^{-3}$, $\xi(T) = [1-(T/T_c)^2]^{1/2}$, and $\chi = d/\lambda(0) \sim 4$.  The model fit the data with parameters $f_{0,g} = 10.80$ GHz, $T_c = 8.47$ K, $\chi = 2.63$, and $\beta = 1.3\times10^{-3}$.  

To fit the $Q$ data in Fig. 1, we used \cite{talanov:2136} $Q^{-1} = Q_b^{-1} + R_\text{eff}/\omega_0 L$, where $Q_b$ is a background $Q$, and the effective surface resistance $R_\text{eff}(T) = R_sG[\coth(d/\lambda) + (d/\lambda)/\sinh^2(d/\lambda)]$, where $G$ is the geometry factor \cite{padamsee}, and a two-fluid model surface resistance was used for $R_s(T)$.  Thus, $Q = \{Q_b^{-1} + [\zeta(T/T_c)^2 (\sinh(2\chi\xi)+2\chi\xi)]/[\xi^2(\xi\sinh^2(\chi\xi)+\beta\sinh(2\chi\xi))]\}^{-1}$, where $\zeta = 2\pi f_0 \lambda^3(0) \sigma \mu_0G/(2R\kappa)$, $f_0 \sim 10$ GHz, and $\sigma$ is the normal-state conductivity.    It is not possible to make an estimate of $\zeta$ because the value of $G$ requires knowledge of the exact current distribution in the SRR.  The data in the left inset of Fig. 1 (red triangles) were fit (black line) using the fit values from $f_0(T)$, giving $Q_b = 6.5\times10^4$, and $\zeta = 1.1\times10^{-5}$.
	
The original hypothesis that Fig. 2 in Ref. [6] was a superposition of many individual SRR resonances is supported by the data in Fig. 1.  We then tested samples with multiple SRRs to see if an equal number of dips (or multiple dips with possibly degenerate frequencies) appear.  A 3-SRR sample was made with a longitudinal lattice parameter of 5 mm, while a 108-SRR array was made by combining four quartz chips, each with a $3\times9$ array of SRRs spaced (center-to-center) 5 mm longitudinally, 3.09 mm vertically, and the chips were spaced $\approx 5$ mm transverse horizontally \cite{ricci:034102}.  Transmission responses of a single (bottom, blue), three (middle, red), and 108-SRR array (top, green) are shown in Fig. 2.  The wide distribution of the 108-SRR array resonant frequencies was presumably due to interactions between the SRRs, and their images in the walls of the waveguide.  Though not shown, it was found that when a single Nb SRR was placed next to a conducting surface, the resonant frequency shifted by 10's of MHz, and the insertion loss and $Q$ decreased.  The data shows that the SRRs are highly interactive, and that the bandwidth of resonant features can increase to over 1 GHz, albeit in a jagged, rather than smooth, manner.  

To test whether the resonant feature in Fig. 1 is due to a $\mu_\text{eff}<0$ feature of the SRR, a resonant transmission peak should be observed when the SRR is embedded in an $\epsilon_\text{eff}<0$ medium.  An array of wires acts as such a medium \cite{pendry:4773}, and so there should be a single transmission peak near the SRR resonant frequency, due to a region of simultaneously negative $\epsilon_\text{eff}$ and $\mu_\text{eff}$, referred to as a negative-index passband.  Nb wire ($T_c$ = 9.25 K by AC susceptibility) with a diameter of 0.25 mm was woven through the perforated waveguide, similar to Ref. [6].  The Nb SRR was placed in the middle of a 5-wide-by-7-long wire array, a distance of 2.29 mm from the center wire, with the capacitive gaps orthogonal to ${\bf E}$, and the wire aligned with the center of the SRR.  Using less than seven rows of wires increased the tunneling background, while increasing the number of rows decreased the height of the transmission peak.

\begin{figure}
\includegraphics[width = 3 in,angle=90]{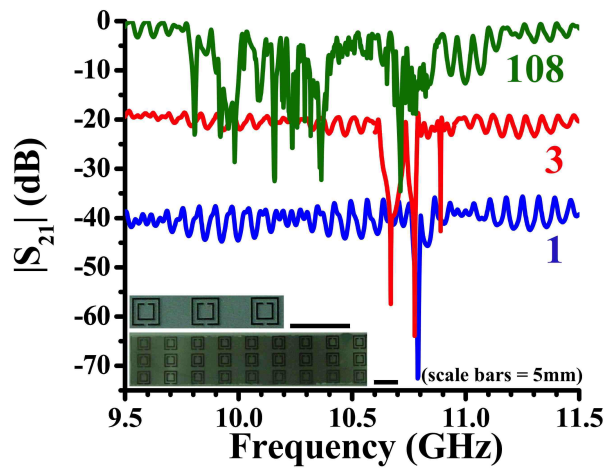}
\caption{Fig. 2 (Color Online) Transmission v. frequency in an empty Nb X-band waveguide for a single (blue, -40 dB offset), three (red, -20 dB offset), and 108 (green) Nb SRR array, all at 5.0 K.  The high frequency ringing in the passband is due to standing wave resonances within the measurement system.  Inset shows 3 and 27-SRR chips.  (Data corrected for cable loss.)}
\label{fig2}
\end{figure}

The transmission results for the SRR embedded in wires are shown in Fig. 3, and revealed an elementary passband near the frequency of the resonant feature in Fig. 1.  The height of the transmission peak decreased and broadened with increasing temperature, and based on the results of the lone Nb SRR, we concluded that the peak was not below the noise floor at temperatures above $T_c$, but disappeared due to the large value of $\rho/t$ in the SRR normal state.  At 4.2 K, we estimated the $Q$ (defined by the 3 dB point below the maximum) to be $2.26\times10^4$.  Replacing the Nb SRR with the Cu-Ni-Au one produced no transmission feature above the noise floor, consistent with the lone SRR transmission results.  In addition, the closed Nb loops also produced no transmission feature above the noise floor.  Further evidence that the features in Fig. 1 were due to $\mu_\text{eff}<0$ comes from comparing the transmission spectra of the lone SRR, and the SRR in the wire array, at the same temperature.  The resonant features occur within 10's of MHz, comparable to the shifts caused by wire-SRR interactions (measured separately).

\begin{figure}
\includegraphics[width=2.7 in,angle=270]{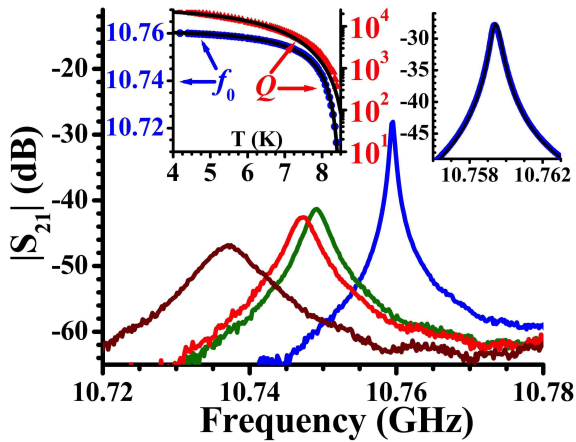}
\caption{Fig. 3 (Color Online) Transmission v. frequency for a single Nb SRR in a $5\times7$ Nb wire array for 5.09 (blue), 7.70 (green), 7.80 (red), and 8.10 (maroon) K (right to left).  Right inset is the 5.09 K peak (blue dots) and fit (black line).  Left inset is the frequency of maximum transmission (blue dots, left axis) and the $Q$ (red triangles, right axis), with fits (black lines) v. temperature.  (Data corrected for cable loss.)}
\label{fig3}
\end{figure}

A single transmission peak at 5.09 K is modeled in the right inset of Fig. 3.  The model used to fit the data was the same as the lone SRR case, but now with two sections of waveguide filled with an $\epsilon_\text{eff} < 0$ medium due to the wires, and the middle section filled with a combination of wires ($\epsilon_\text{eff} < 0$) and an SRR ($\mu_\text{eff} < 0$) to give $n_\text{eff} < 0$.  Fixing the lengths of the sections, we found $f_c = 18$ GHz, $\gamma/2\pi = 500$ kHz, $f_0 = 10.72$ GHz, $\Gamma/2\pi = 600$ kHz, and $F = 0.032$.  Here, $f_c$ is the cutoff frequency of the wire array and $\gamma$ is the loss parameter  \cite{pendry:4773} of $\epsilon_\text{eff}$, and these fit values were consistent with data obtained in Ref. [6].  Note that $\Gamma$ has increased, while $F$ has decreased, relative to lone SRR results, signs of the wire-SRR interaction.    

Similar to the lone SRR data, the negative-index feature changed with temperature, as the kinetic inductance increased (left inset, Fig. 3).  The frequency shift was modeled as before, and the fit parameter results were similar; $f_{0,g} = 10.77$ GHz, $T_c = 8.67$ K, and $\beta = 1.5\times10^{-3}$, while fixing $\chi = 2.63$.  The difference in $f_{0,g}$ between the lone and embedded SRRs can be attributed to a wire-SRR interaction, similar to the multiple SRR case.  The $Q$ data were fit by $Q_b = 2.7\times10^4$, and $\zeta = 1.2\times10^{-5}$, and these values are also consistent with the lone SRR.
	 
	The transmission response of a single SRR was observed in an empty all-Nb waveguide, and in a Nb wire medium.  In the empty waveguide, the SRR produced a very high-$Q$ transmission dip due to a frequency range of $\mu_\text{eff} < 0$, while the SRR in the wire array produced a sharp transmission peak due to $n_\text{eff} < 0$.  Both the lone and embedded SRR transmission features disappeared above $T_c$, due to the thin-film geometry and losses.  A single thick Cu-Ni-Au SRR produced no transmission features, indicating that superconducting metals and low-loss dielectrics are better suited for scaled down (and few) SRR applications in the microwave regime.  Transmission features were temperature dependent, and fit with simple models of superconductor electrodynamics.  The data shows that tunable filters and passbands with very high $Q$'s can be created, and with appropriate engineering, they can cover a large range of frequencies, though care is needed to account for perturbations when using multiple SRRs.

	This work was supported by the NSF through Grant No. NSF/ECS-0322844, and the DI Outreach Program.  We would like to thank J. Hamilton and N. Orloff for assistance, and P. Kneisel and L. Turlington  at TJNAL for their help in making the Nb waveguides and couplers.






\newpage
\begin{thebibliography}{99}

\bibitem{veselago:517} V. G. Veselago, Usp. Fiz. Nauk {\bf 92}, 517 (1967) [English translation: Sov. Phys. Usp. {\bf 10}, 509 (1968)].

\bibitem{shelby:77} R. A. Shelby, D. R. Smith, and S. Schultz, Science {\bf 292}, 77 (2001).

\bibitem{bayindir:120} M. Bayindir, K. Aydin, E. Ozbay, P. Markos, and C. M. Soukoulis, Appl. Phys. Lett. {\bf 81}, 120 (2002).

\bibitem{houck:137401} A. A. Houck, J. B. Brock, and I. L. Chuang, Phys. Rev. Lett. {\bf 90}, 137401 (2003).

\bibitem{cubukcu:604} E. Cubukcu, K. Aydin, E. Ozbay, S. Foteinopoulou, and C.M. Soukoulis, Nature (London) {\bf 423}, 604 (2003).

\bibitem{ricci:034102} M. Ricci, N. Orloff, and S. M. Anlage, Appl. Phys. Lett. {\bf 87}, 34102 (2005).

\bibitem{pendry:3966} J. B. Pendry, Phys. Rev. Lett. {\bf 85}, 3966 (2000).

\bibitem{shen:3286} J. T. Shen and P. M. Platzman, Appl. Phys. Lett. {\bf 80}, 3286 (2002).

\bibitem{gomez-santos:077401}G. Gomez-Santos, Phys. Rev. Lett. {\bf 90}, 77401 (2003).

\bibitem{smith:1506} D. R. Smith, D. Schurig, M. Rosenbluth, S. Schultz, S. A. Ramakrishna, and J. B.  Pendry, Appl. Phys. Lett. {\bf 82}, 1506 (2003).

\bibitem{ziolkowski:2626}R. W. Ziolkowski and A. D. Kipple, IEEE Trans. Antennas and Propagation {\bf 51}, 2626 (2003).

\bibitem{dimmock:2397} J. O. Dimmock, Optics Express {\bf 11}, 2397 (2003).

\bibitem{zhou:223902} J. Zhou, T. Koschny, M. Kafesaki, E. N. Economou, J. B. Pendry, and C. M. Soukoulis, Phys. Rev. Lett. {\bf 95}, 223902 (2005).

\bibitem{pendry:4773} J. B. Pendry, A. J. Holden, W. J. Stewart, and I. Youngs, Phys. Rev. Lett. {\bf 76}, 4773 (1996).

\bibitem{koschny:107402} T. Koschny, M. Kafesaki, E. N. Economou, and C. M. Soukoulis, Phys. Rev. Lett. {\bf 93}, 107402 (2004).

\bibitem{padilla:0508307} W. J. Padilla, cond-mat/0508307.

\bibitem{marques:144440} R. Marqu\'es, F. Medina, and R. Rafii-El-Idrissi, Phys. Rev. B {\bf 65}, 144440 (2002).

\bibitem{gay-balmaz:2929} P. Gay-Balmaz and O. J. F. Martin, J. Appl. Phys. {\bf 92}, 2929 (2002).

\bibitem{baena:75116} J. D. Baena, L. Jelinek, R. Marqu\'es, and F. Medina, Phys. Rev. B {\bf 72}, 75116 (2005).

\bibitem{pendry:2075} J. B. Pendry, A. J. Holden, D. J. Robbins, and W. J. Stewart, IEEE Trans. Microwave Theory Tech. {\bf 47}, 2075 (1999).

\bibitem{anlage:1388} S. M. Anlage, H. J. Snortland, and M. R. Beasley, IEEE Trans. Magnetics {\bf 25}, 1388 (1989).

\bibitem{hoffmann} T. K. Ishii, ed., {\it Handbook of Microwave Technology}, vol. 1 (Academic Press, Inc., San Diego, CA, 1995), chp. 4.

\bibitem{henkels:63} W. H. Henkels and C. J. Kircher, IEEE Trans. Magnetics {\bf 13}, 63 (1977).

\bibitem{bluzer:2051} N. Bluzer and D. K. Fork, IEEE Trans. Magnetics {\bf 28}, 2051 (1992).

\bibitem{talanov:2136} V. V. Talanov, L. V. Mercaldo, S. M. Anlage, and J. H. Claassen, Rev. Sci. Instrum. {\bf 71}, 2136 (2000).

\bibitem{padamsee} H. Padamsee, J. Knobloch, and T. Hays, {\it RF Superconductivity for Accelerators} (John Wiley \& Sons, Inc., New York, NY, 1998), chp. 2.



\end{thebibliography}
 \end{document}